\documentclass[10pt]{article}
\usepackage{multicol}
\usepackage{graphicx}
\usepackage{amsmath}
\usepackage[a4paper]{geometry}
\usepackage{hyperref}
\usepackage{rotating}

\begin{document}

\begin{center}
{\Large \bf An evidence of triple kinetic freezeout scenario observed in all centrality intervals in Cu-Cu, Au-Au and Pb-Pb collisions at high energies}

\vskip1.0cm

M.~Waqas$^{1,}${\footnote{Corresponding author. Email (M.Waqas):
waqas\_phy313@yahoo.com; waqas\_phy313@ucas.ac.cn}},G. X. Peng$^{1,2}$ {\footnote{Corresponding author. Email (G. X. Peng): gxpeng@ucas.ac.cn}},
Fu-Hu Liu$^{3,4}${\footnote{E-mail: fuhuliu@163.com; fuhuliu@sxu.edu.cn}}
\\

{\small\it  $^1$ School of Nuclear Science and Technology, University of Chinese Academy of Sciences,
Beijing 100049, China,

$^2$ Theoretical Physics Center for Science Facilities, Institute of High Energy Physics, Beijing 100049, China,

$^3$Institute of Theoretical Physics \& State Key
Laboratory of Quantum Optics and Quantum Optics Devices,\\ Shanxi
University, Taiyuan, Shanxi 030006, China

$^4$Collaborative Innovation Center of Extreme Optics, Shanxi University, Taiyuan, Shanxi 030006, China}

\end{center}
Note: All authors are equally contributed.
\vskip1.0cm

{\bf Abstract:} Transverse momentum spectra of $\pi^+$, $K^+$, $p$, $K^0_s$, $\Lambda$, $\Xi$ or $\bar\Xi^+$ and $\Omega$ or
$\bar\Omega^+$ or $\Omega+\bar\Omega$ in Copper-Copper (Cu-Cu), Gold-Gold (Au-Au)
and Lead-Lead (Pb-Pb) collisions at 200 GeV, 62.4 GeV and 2.76 TeV respectively, are analyzed in different centrality bins by the blast wave model with Tsallis
statistics. The model results are approximately in agreement with the experimental data measured by BRAHMS, STAR and ALICE Collaborations in special transverse momentum ranges. Kinetic freeze out temperature, transverse flow velocity and kinetic freezeout volume are extracted from the transverse momentum spectra of the particles.
It is observed that $\bar\Xi^+$ and $\Omega$ or $\bar\Omega^+$ or $\Omega+\bar\Omega$ have larger kinetic freezeout temperature followed by $K^+$, $K^0_s$ and $\Lambda$ than $\pi^+$ and $p$ due to smaller reaction cross-sections of multi-strange and strange particles than non-strange particles. The present work reveals the scenario of triple kinetic freezeout in collisions at BRAHMS, STAR and ALICE Collaborations, however the transverse flow velocity and kinetic freezeout volume are mass dependent and they decrease with the increasing rest mass of the particle. In addition, the kinetic freezeout temperature, transverse flow velocity and kinetic freezeout volume are decreasing from central to peripheral collisions while the parameter q increase from central to peripheral collisions, indicating the approach of quick equilibrium in the central collisions. Besides, the kinetic freezeout temperature and kinetic freezeout volume are observed to be larger in larger collision system which shows its dependence on the size of the interacting system, while transverse flow velocity increase with increasing energy.
\\

{\bf Keywords:} non-strange, strange, multi-strange, kinetic freeze-out temperature, transverse flow
velocity, kinetic freezeout volume, cross-section, centrality bins, transverse momentum spectra.

{\bf PACS:} 12.40.Ee, 13.85.Hd, 25.75.Ag, 25.75.Dw, 24.10.Pa

\vskip1.0cm

\begin{multicols}{2}

{\section{Introduction}}
One of the newest trends in the advancement of relativistic heavy ion collisions is to dig out the new states
of strongly interacting matter and to ascertain the Quark Gluon Plasma (QGP) [1--4] anticipated qualitatively
by the Quantum Chromodynamics (QCD) [5--10]. It is considered that the QGP is a state of strongly interacting
matter under the extreme conditions of high temperatures and/or high net baryon densities. Nowadays, the QGP can be
created in the laboratories during the Gold-Gold (Au-Au) and Copper-Copper (Cu-Cu) collisions at Relativistic
Heavy Ion Collider (RHIC), Lead-Lead (Pb-Pb) and Xenon-Xenon (Xe-Xe) collisions at Large Hadron Collider (LHC)
and other high energy nucleus-nucleus collisions by increasing the energies and altering the masses of the
colliding nuclei.
In fact, an extremely high temperature and/or high densities are required for the formation of QGP. In the present work,
we will be limited to the concept of temperature. Indeed temperature is a very important concept in thermal and
sub-atomic particles [11] due to its wide applications in experimental and theoretical studies. In literature, one can
find various kinds of temperatures at various stages of the collision. The degree of excitation of interacting system
at the initial stage of collisions is described by the initial temperature. The chemical freezeout temperature
describes the excitation degree of chemical freezeout,when the inelastic scattering cease and the particle identities
are set until decay [12, 13], followed by the kinetic freezeout temperature which describes the degree of excitation
at the kinetic freezeout stage.  Since, the present work is mainly focused on the behaviour of kinetic freezeout temperature,
therefore we will not discuss the other temperatures here, but one can have a look  at [14, 15] for their more details.
Generally, the freezeout could be very complicated process due to the involvement of duration in time and a hierarchy
where various kinds of particles and reactions switch-off at different times. From kinematic point of view, the reactions
with lower cross-section is expected to be switched-off at higher densities/temperatures or early in time than the
reactions with higher cross-sections. Hence, the chemical freezeout occur earlier in time than the kinetic freezeout,
which correspond to the elastic reactions. According to the above statement, it is believed that the charm and strange
particles decouple from the system earlier than the lighter hadrons. A series of freezeouts maybe possibly correspond
to particular reaction channels [16]. Furthermore, a single [17], double [13, 18] and multiple kinetic freezeout scenarios
[19--21] can be found in literature. In addition, the transverse flow velocity ($\beta_T$) is also an important parameter which reflects the collective expansion of the emission source. Both $T_0$ and $\beta_T$ can be extracted from the transverse momentum ($p_T$) spectra of the particles by using some distribution laws.

Additionally, volume keeps a prominent importance in collision physics. The volume occupied by the sources of ejectiles
when the mutual interactions become negligible and they only feel the coulombic repulsive force is known as kinetic freezeout volume ($V$).
Various freezeout volume correspond to various stages in interaction process but we will only focus on the kinetic
freezeout volume in the present work.

The $p_T$ spectra of the particles produced in high energy collisions are very important quantities to be measured.
The study of $p_T$ spectra of particles can give the understanding of the production mechanism and help us to infer the thermodynamic
properties of the collision system. Beside, it gives some useful information that contains, but not limited to various types of
temperatures and freezeout volumes.

In the present work, we analyzed the $p_T$ spectra of $\pi^+$, $K^+$, $p$, $K^0_s$, $\Lambda$, $\Xi$ or $\bar\Xi^+$ and $\Omega$ or
$\bar \Omega^+$ or $\Omega+\bar\Omega$ by using the blast wave model with Tsallis statistics and extracted the  parameters $T_0$, $\beta_T$ and $V$.

The remainder of the paper consists of method and formalism in section 2, followed by the results
and discussion in section 3. In section 4, we summarized our main observations and conclusions.
\\

{\section{The method and formalism}} The structure of $p_T$ spectra in high energy collisions
are complex processes in which many emission sources
are found. A local equilibrium state may possibly be formed by the sources with the same excitation degree, which can be described by the standard distribution. In case of different equilibrium states with different degrees of excitation, we may use different parameters of temperature. In general, the $p_T$ spectra in a not too wide $p_T$ range can be described by the two or three components standard distribution that reflects the fluctuation of temperature of the interacting system. Meanwhile, a two- or three-components standard distribution can be described by the Tsallis distribution [22--26], blast wave model with Tsallis (TBW) statistics [27--29] or others.

According to ref [27--29], the TBW model results in the $p_T$ distribution to be
\begin{align}
f_1(p_T)=&\frac{1}{N}\frac{\mathrm{d}N}{\mathrm{d}p_\mathrm{T}}=C \frac{gV}{(2\pi)^2} p_T m_T \int_{-\pi}^\pi d\phi\int_0^R rdr \nonumber\\
& \times\bigg\{{1+\frac{q-1}{T_0}} \bigg[m_T  \cosh(\rho)-p_T \sinh(\rho) \nonumber\\
& \times\cos(\phi)\bigg]\bigg\}^\frac{-1}{(q-1)}
\end{align}
where $C$ stands for the normalization constant that leads the integral in Eq. (1) to be normalized to 1, $g$ is the degeneracy factor which is different for different particles based on $g_n$=2$S_n$+1 ($S_n$ is the spin of the particle), $m_T=\sqrt{p_T^2+m_0^2}$
is the transverse mass, $m_0$ is the rest mass of the particle, $\phi$ represents the azimuthal angle,
r is the radial coordinate, $V$ is the freezeout volume,  $T_0$ is the kinetic freezeout temperature, $R$ is the maximum $r$, $q$ is the entropy index and it shows the measure of degree of deviation of the system from an equilibrium state,
$\rho=\tanh^{-1} [\beta(r)]$ is the boost angle, $\beta(r)=\beta_S(r/R)^{n_0}$
is a self-similar flow profile, $\beta_S$ is the flow velocity on the surface,as mean of $\beta(r)$,$\beta_T=(2/R^2)\int_0^R r\beta(r)dr=2\beta_S/(n_0+2)=2\beta_S/3$,
and $n_0$ =1 or 2 [29]. $n_0$ is regarded a free parameter in some literature [30, 31]. In case we choose $n_0$ as a free parameter, we will have one more free parameter, which is not chosen by us. Furthermore, the index $-1/(q-1)$ in Eq.(1) can be replaced by $-q/(q-1)$ due to the reason that $q$ is being close to 1. This replacement results in a small and negligible divergence in the Tsallis distribution [22,32].

If the $p_T$ spectra is not too wide, it can be described by Eq.(1) and $T_0$ and $\beta_T$ can be extracted, but in case of wide $p_T$ spectra the contribution of hard scattering process will be considered and according to the QCD calculus [33--35] it can be parameterized as an inverse power law.
\begin{align}
f_H(p_T)=&\frac{1}{N}\frac{\mathrm{d}N}{\mathrm{d}p_\mathrm{T}}= Ap_T \bigg( 1+\frac{p_T}{p_0} \bigg)^{-n},
\end{align}
which is the Hagedorn function, $A$ is the normalization constant, $p_0$ and $n$ are the free parameters. The modified versions of Hagedorn function can be found in literature [36--42].
\begin{align}
f_H(p_T)=&\frac{1}{N}\frac{\mathrm{d}N}{\mathrm{d}p_\mathrm{T}}=\frac{Ap^2_T}{m_T}\bigg( 1+\frac{p_T}{p_0} \bigg)^{-n},
\end{align}
\begin{align}
f_H(p_T)=&\frac{1}{N}\frac{\mathrm{d}N}{\mathrm{d}p_\mathrm{T}}= Ap_T \bigg( 1+\frac{p^2_T}{p^2_0} \bigg)^{-n},
\end{align}
\begin{align}
f_H(p_T)=&\frac{1}{N}\frac{\mathrm{d}N}{\mathrm{d}p_\mathrm{T}}= A\bigg( 1+\frac{p^2_T}{p^2_0} \bigg)^{-n},
\end{align}
where as in Eq.(3),(4) and (5), $p_0$ and $n$ are severally different.
In case of wide $p_T$ range, the superposition of soft excitation and hard scattering process can be used for its description. If eqn.(1) describe the soft excitation process, and the hard scattering process is described by one of Eq.(2)-(5), then the superposition of the two-components can be used to describe the wide $p_T$ range.
\begin{align}
f_0(p_T)=kf_S(p_T)+(1-k)f_H(p_T),
\end{align}
According to Hagedorn model [43], the usual step function can also be used for the superposition of the two functions, as
\begin{align}
f_0(p_T)=A_1\theta(p_1-p_T) f_S(p_T) + A_2 \theta(p_T-p_1)f_H(p_T),
\end{align}
where $f_S$ and $f_H$ are the soft and hard components respectively,  $k (1-k)$ shows the contribution fraction of soft excitation (hard scattering)
process which naturally normalizes to 1. $A_1$ and $A_2$ are the normalization constants that synthesize $A_1$$f_S$$(p_1)$=$A_1$$f_H$$(p_1)$ and $\theta$(x) is the usual step function. The soft component $f_S(p_T)$ in Eqn.6 and Eqn.7 are the same as $f_1(p_T)$ in Eqn. (1).

The contribution of soft component in low $p_T$ and hard component in high $p_T$ in Eq.(7) are linked with each other at $p_T$=$p_1$.

 The soft and hard components in Eq.(6) and (7) are treated in different ways in the whole $p_T$ region. Eq.(6) refers to the contribution of soft component in the range 0-2$\sim$3 GeV/c or a little more. However in case of the contribution of hard component, though the main contribution in low $p_T$ region is soft excitation process, but it covers the whole $p_T$ region. In Eq.(7), in the range from 0 to $p_1$ and from $p_1$ up to maximum are the contributions of soft and hard components respectively and there is no mixed region for the two components.

 Eq.(6) and (7) are the same if only soft component is included. However if we include both the soft and hard components, small variation in parameters will appear.
\\

{\section{Results and discussion}}
Figure 1 presents the event centrality dependent double
differential $m_T$ or $p_T$ spectra, $1/N_{ev}$[(1/2$\pi$$m_T$) $d^2$$N$/$dyd$$m_T$] of
$\pi^+$,$K^+$ and $p$ and [(1/2$\pi$$p_T$) $d^2$$N$/$dyd$$p_T$] of $K^0_S$ produced in
Cu-Cu collisions at the center-of-mass energy per nucleon pair
$\sqrt{s_{NN}}=200$ GeV in panels (a)-(d) respectively. The symbols represent the experimental data measured by
the BRAHMS and STAR Collaborations [44, 45] and the curves are our fitting
results by using the blast-wave model with Tsallis statistics, Eq. (1). The spectra is distributed in different centrality classes,
e:g for $\pi^+$,$K^+$ and $p$ 0--10\%, 10--30\%, 30--50\% and 50--70\% $|y|=0$, while for
$K^0_S$ 0--10\%, 10--20\%, 20--30\%,30--40\%  and 40--60\% $|y|<0.5$, where $y$ denotes the
rapidity. The corresponding ratio of data/fit is followed in each panel. The related parameters along with $\chi^2$ and degree of freedom (dof) are listed in Table 1, where the centrality classes and the re-scaled spectra are also presented together. The data is chosen in order to search the differences in different particles emission. One can see that Eq. (1) fits well the data in Cu-Cu collisions at 200 GeV at the RHIC. It is noticeable that in the present work, we have analyzed the $p_T$ distribution in a limited interval and considered the soft excitation process and so parametrization (1) is used. If the $p_T$ range becomes wider then hard scattering will be involved and the parametrization of Eq. (2)-(7) can be used.

The normalization constant $N_0$ is
used for comparison of the fit function $fS(p_T)$ or $fS(m_T)$ and
the experimental spectra, and the normalization constant $C$
is used to let the integral of Eq. (1) be unity. The two normalization
constants are not the same, though $C$ can be absorbed
in $N_0$. We have used both $C$ and $N_0$ to present a clear description.
\begin{figure*}[htb!]
\begin{center}
\hskip-0.153cm
\includegraphics[width=15cm]{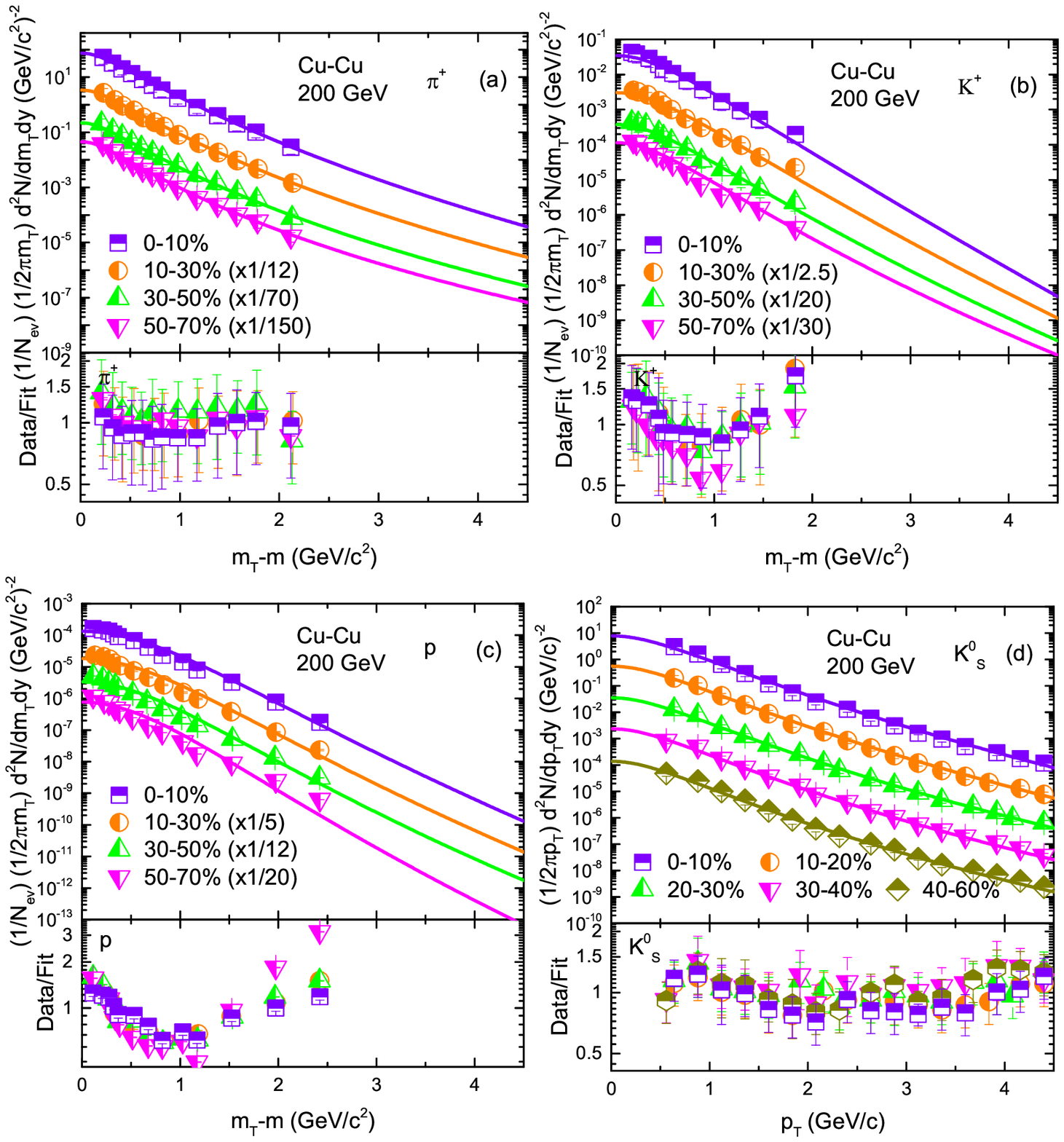}
\end{center}
Fig. 1. Transverse mass spectra of (a)-(d) $\pi^+$, $K^+$,
$p$ and $K^0_S$ produced in different centrality bins in Cu-Cu collisions at
$\sqrt{s_{NN}}=200$ GeV. The symbols represent the experimental
data measured by the BRAHMS Collaboration at
$|y|=0$ [44] for $\pi^+$, $K^+$ and $p$ and by STAR Collaboration $|y|<0.5$ in [45] for $K^0_S$. The curves are our fitted results by
Eq. (1). Each panel is followed by its corresponding ratios of Data/Fit.
\end{figure*}
\begin{figure*}[htb!]
\begin{center}
\hskip-0.153cm
\includegraphics[width=15cm]{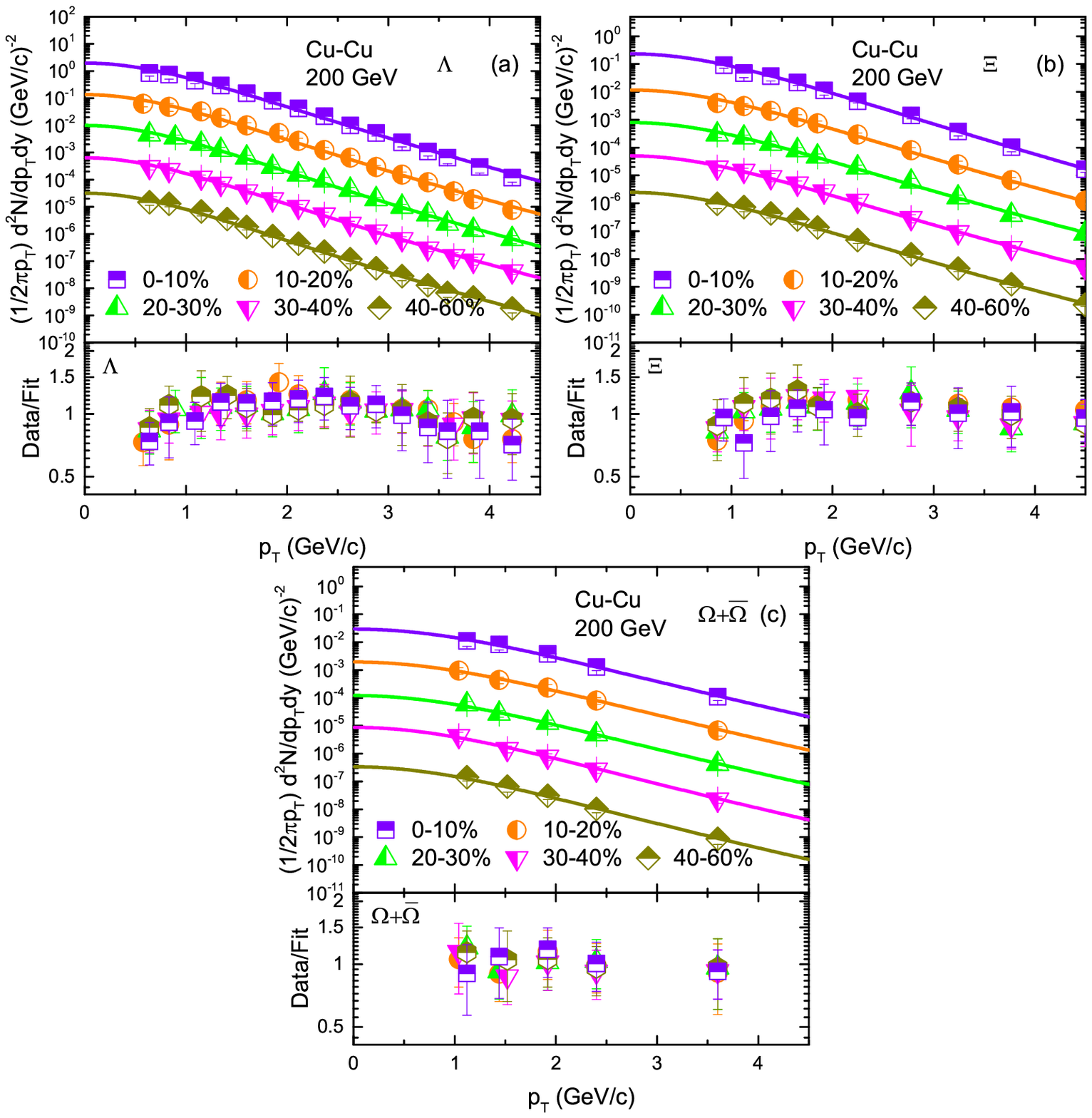}
\end{center}
Fig. 2. Transverse momentum spectra of (a)-(c) $\Lambda$, $\Xi$ and
$\Omega+\bar\Omega$ produced in different centrality bins in Cu-Cu collisions at
$\sqrt{s_{NN}}=200$ GeV. The symbols represent the experimental
data measured by the STAR Collaboration at rapidity
$|y|<0.5$ [44]. The curves are our fitted results by
Eq. (1). Each panel is followed by its corresponding ratios of Data/Fit.
\end{figure*}

Figure 2 is similar to Figure 1, but it shows the $p_T$ spectra of $\Lambda$, $\Xi$ and $\Omega+\bar \Omega$ in panels (a)-(c) with $|y|<0.5$ in
Cu-Cu collisions at 200 GeV. The spectra of $\Lambda$, $\Xi$ and $\Omega+\bar \Omega$ is distributed in different centrality classes of 0--10\%, 10--20\%, 20--30\%, 30--40\% and 40--60\%. The symbols represent the experimental data of STAR [44] collaboration while the curves are the results of our fitting by using Eq. (1). Each panel is followed by result of its data/fit. One can see that the model results describe approximately the experimental data in special $p_T$ ranges at RHIC.
\begin{figure*}[htb!]
\begin{center}
\hskip-0.153cm
\includegraphics[width=15cm]{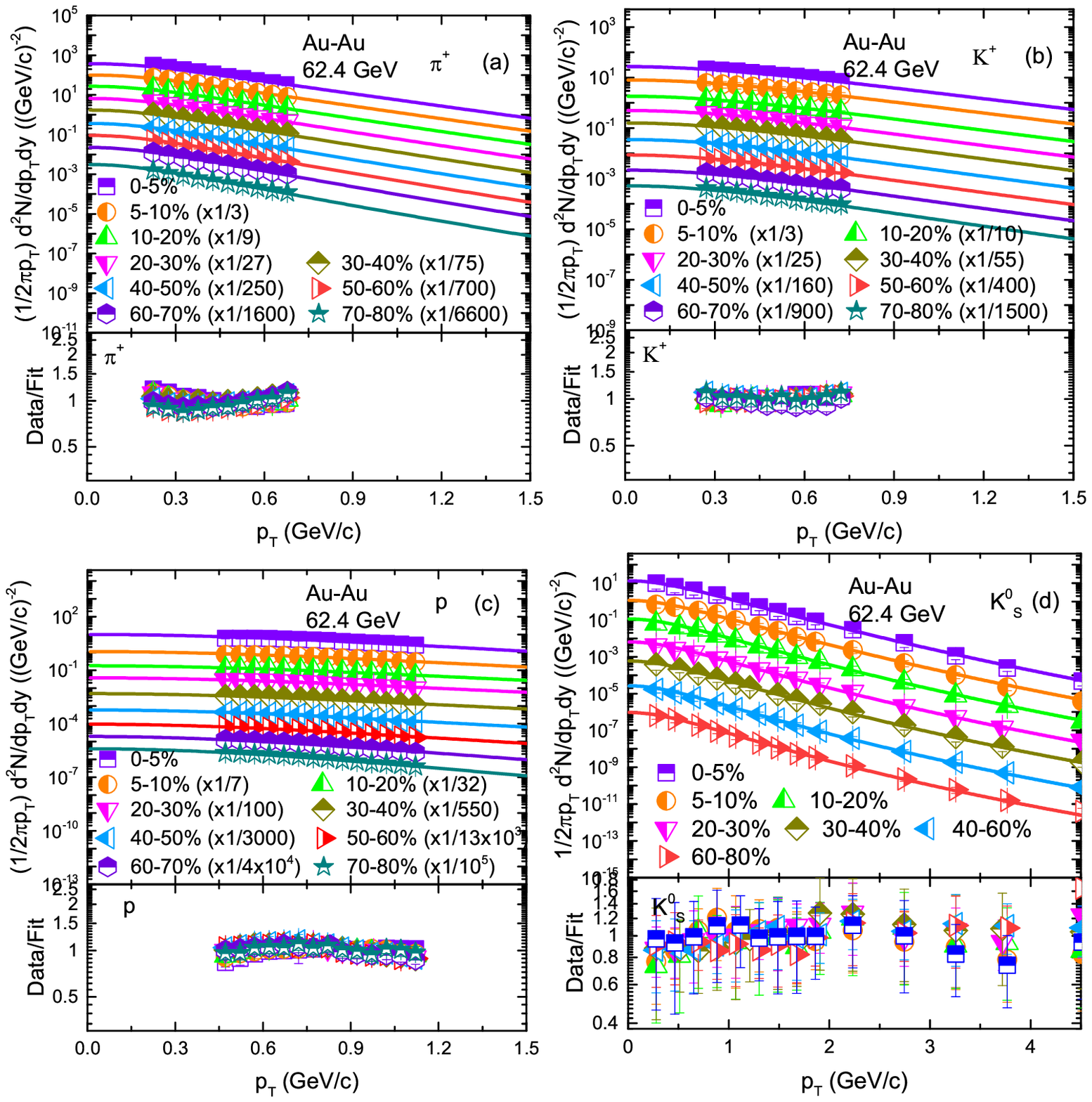}
\end{center}
Fig. 3. Transverse mass spectra of (a)-(d) $\pi^+$, $K^+$
$p$ at rapidity $|y|<0.1$ [31] and $K^0_S$ at $|\eta| < 1.8$ [46] produced in different centrality bins in Au-Au collisions at
$\sqrt{s_{NN}}=62.4$ GeV. The symbols represent the experimental data measured by the STAR Collaboration. The curves are our fitted results by
Eq. (1). Each panel is followed by its corresponding ratios of Data/Fit.
\end{figure*}

Figures 3 and 4 are similar to Figure 1, but it shows the $p_T$ spectra of
$\pi^+$, $K^+$, $p$ and $K^0_S$ in fig3.(a)-3(d)  and $\Lambda$, $\bar \Xi^+$ and $\bar\Omega^+$ in fig4.(a)-4(c) in Au-Au collisions at 62.4 GeV. The spectra is distributed in different centrality classes, e.g for $\pi^+$, $K^+$ and $p$ 0--5\%, 5--10\%, 10--20\%, 20--30\%, 30--40\%, 40--50\%, 50--60\%, 60--70\% and 70--80\% , while for $K^0_S$, $\Lambda$, and $\bar \Xi^+$ 0--5\%, 5--10\%, 10--20\%, 20--30\%, 30--40\%, 40--60\% and 60--80\%, and for and $\bar\Omega^+$ is 0--20\%, 20--40\% and 40--60\%. The symbols represent the experimental data of STAR Collaboration [31, 46], while the curves are the results of our fitting by using Eq. (1). Each panel is followed by result of its data/fit. One can see that the model results describe approximately the experimental data in special $p_T$ ranges at RHIC. The special $p_T$ range refers to the contribution of soft excitation process. Furthermore the data for $K^0_S$, $\Lambda$, $\Xi$ and $\Omega+\bar \Omega$ in Cu-Cu collisions and  $K^0_S$ in Au-Au collision is cut at higher at high $p_T$.
\begin{figure*}[htb!]
\begin{center}
\hskip-0.153cm
\includegraphics[width=15cm]{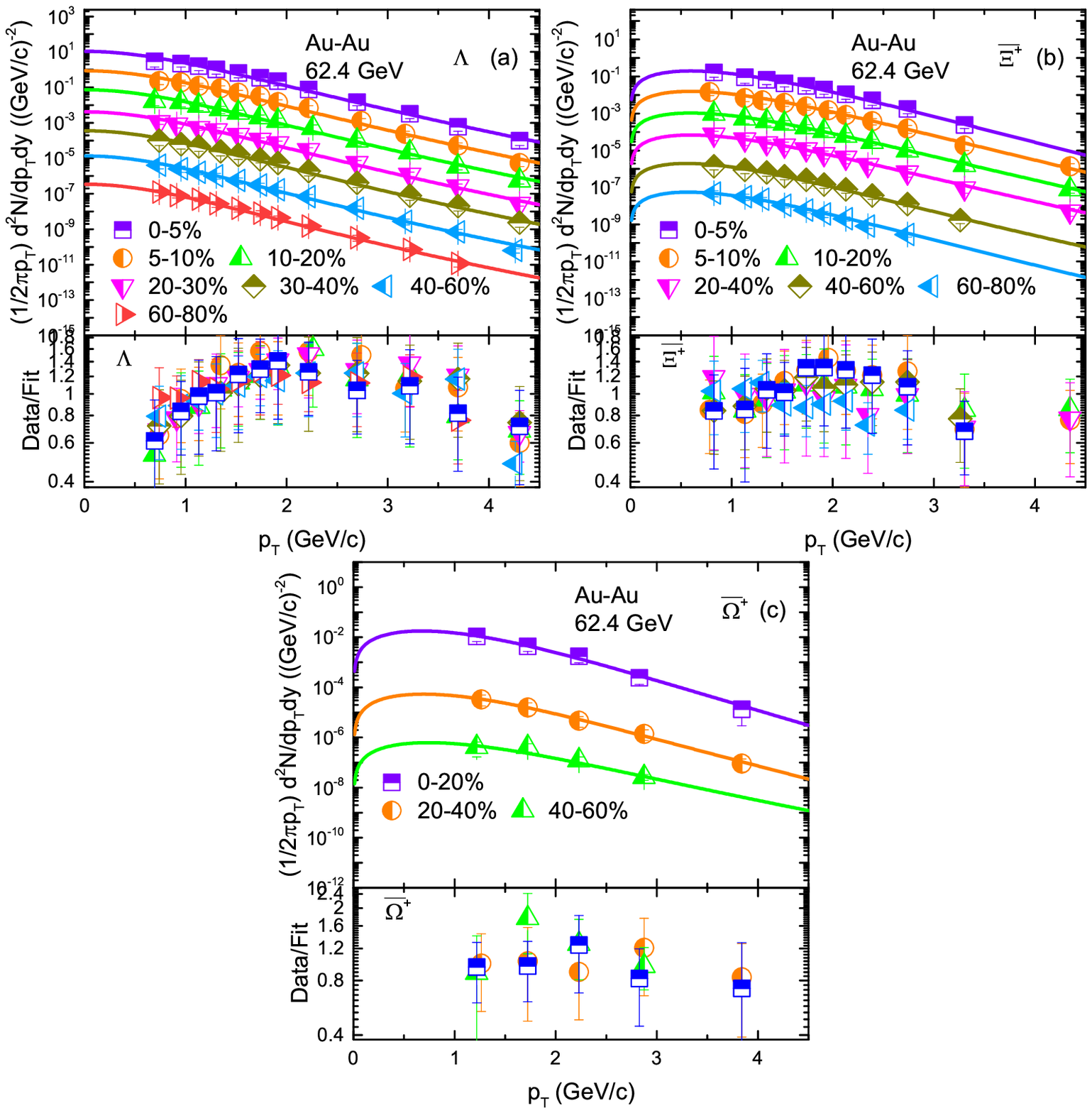}
\end{center}
Fig. 4. Transverse momentum spectra of (a)-(d) $\Lambda$, $\bar \Xi^+$ and
$\bar\Omega^+$ produced in different centrality bins in Au-Au collisions at
$\sqrt{s_{NN}}=62.4$ GeV. The symbols represent the experimental
data measured by the STAR Collaboration at rapidity
$|\eta| < 1.8$ [46]. The curves are our fitted results by
Eq. (1). Each panel is followed by its corresponding ratios of Data/Fit.
\end{figure*}
\begin{figure*}[htb!]
\begin{center}
\hskip-0.153cm
\includegraphics[width=15cm]{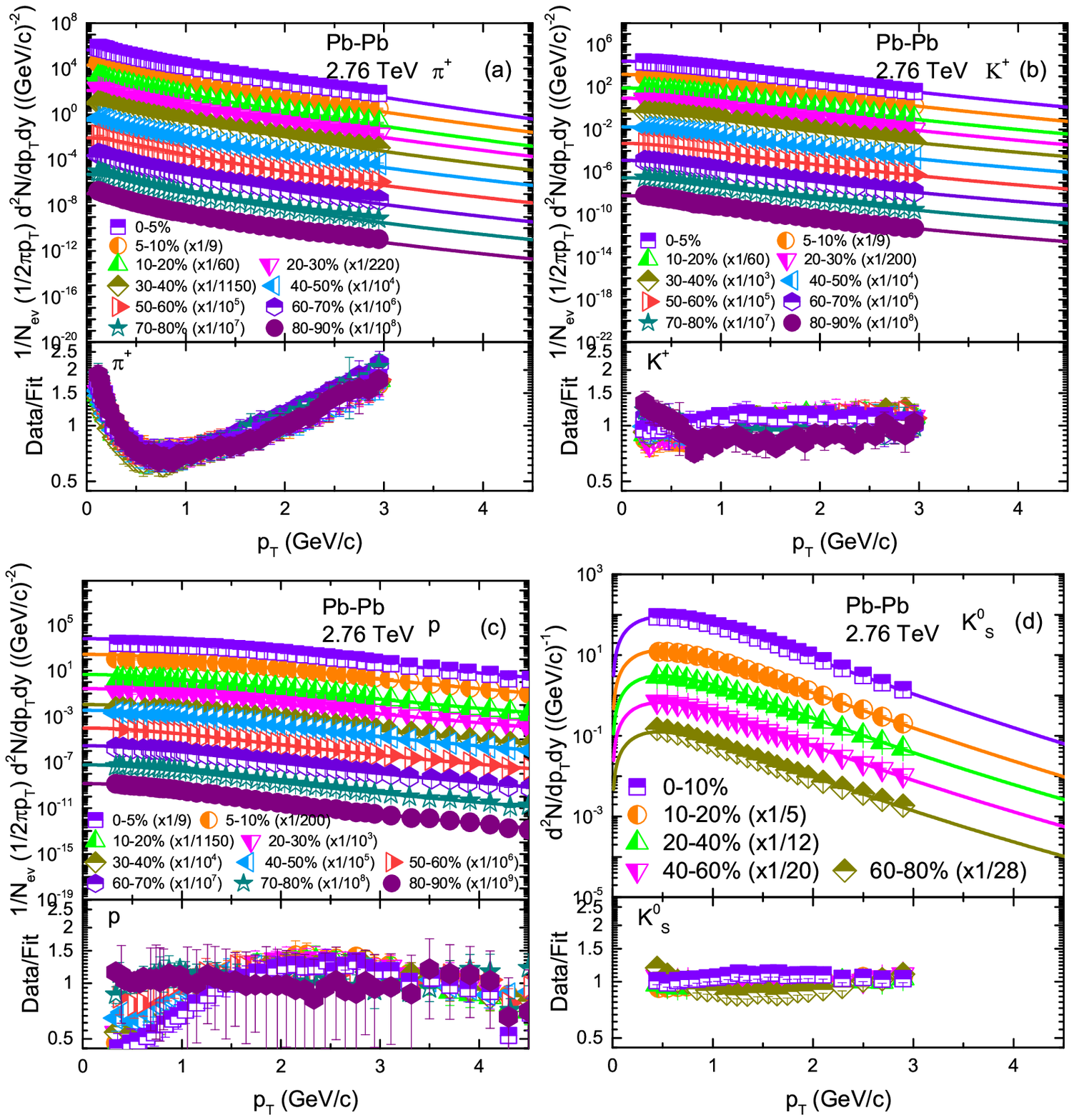}
\end{center}
Fig. 5. Transverse momentum spectra of (a)-(d) $\pi^+$, $K^+$
$p$ and $K^0_S$ produced in different centrality bins in Pb-Pb collisions at
$\sqrt{s_{NN}}=2.76$ TeV. The symbols represent the experimental
data measured by the ALICE Collaboration at rapidity
$|y|<0.5$ [47, 48]. The curves are our fitted results by
Eq. (1). Each panel is followed by its corresponding ratios of Data/Fit.
\end{figure*}
\begin{figure*}[htb!]
\begin{center}
\hskip-0.153cm
\includegraphics[width=15cm]{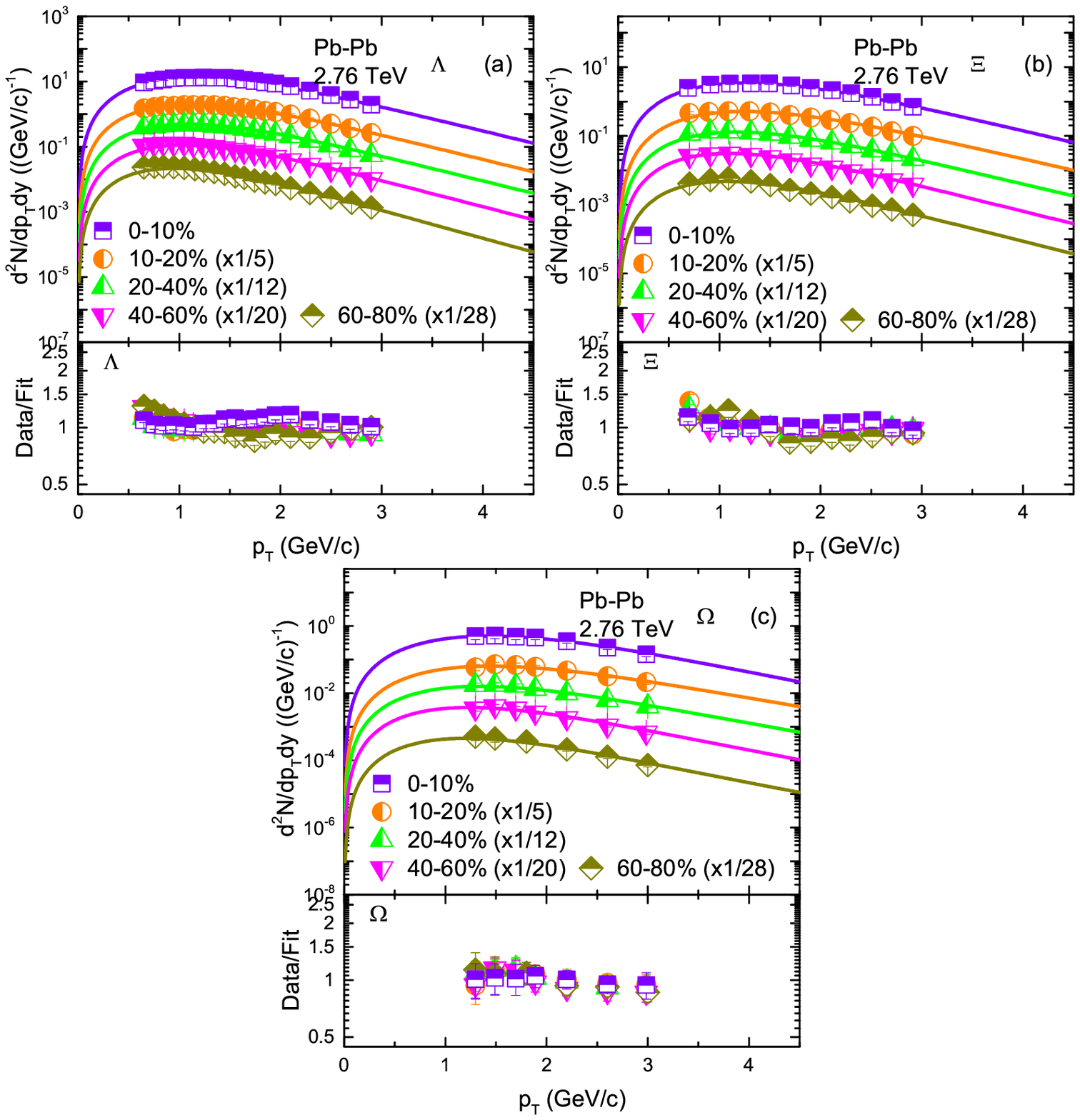}
\end{center}
Fig. 6. Transverse momentum spectra of (a)-(c) $\Lambda$, $\Xi$ and
$\Omega$ produced in different centrality bins in Pb-Pb collisions at
$\sqrt{s_{NN}}=2.76$ TeV. The symbols represent the experimental
data measured by the ALICE Collaboration at rapidity $|y|<0.5$ [48]. The curves are our fitted results by
Eq. (1). Each panel is followed by its corresponding ratios of Data/Fit.
\end{figure*}
\begin{table*}
{\scriptsize Table 1. Values of free parameters $T_0$ and
$\beta_T$, V and q, normalization constant ($N_0$),
$\chi^2$, and degree of freedom (dof) corresponding to the curves
in Figs. 1--6. \vspace{-.50cm}
\begin{center}
\begin{tabular}{ccccccccccc}\\ \hline\hline
Collisions       & Centrality       & Particle   & $T_0$ (GeV)        & $\beta_T$ (c)       & $V (fm^3)$   & $q$        & $N_0$   & $\chi^2$/ dof \\ \hline
Fig. 1           & 0--10\%          & $\pi^+$     &$0.104\pm0.005$  & $0.389\pm0.010$  & $4000\pm270$ & $1.05\pm0.005$   &$0.07\pm0.003$  & 1/9\\
Cu-Cu            & 10--30\%         & --         &$0.102\pm0.005$  & $0.380\pm0.009$  & $3800\pm210$ & $1.06\pm0.004$   &$0.04\pm0.005$  & 0.6/9\\
200 GeV          & 30--50\%         &--          &$0.100\pm0.005$  & $0.360\pm0.010$  & $3700\pm200$ & $1.07\pm0.006$   &$0.015\pm0.004$ & 2/9\\
                 & 50--70\%         & --         &$0.097\pm0.005$  & $0.340\pm0.010$  & $3558\pm220$ & $1.08\pm0.008$   &$0.0065\pm0.0006$& 1.5/9\\
\hline
                 & 0--10\%          & $K^+$      &$0.128\pm0.007$  & $0.360\pm0.013$  & $3570\pm177$ & $1.01\pm0.005$   &$0.00007\pm0.000004$& 2/8\\
                 & 10--30\%         & --         &$0.126\pm0.005$  & $0.353\pm0.007$  & $3490\pm220$ & $1.02\pm0.006$   &$0.000016\pm0.000006$& 2/8\\
                 & 30--50\%         &--          &$0.124\pm0.005$  & $0.335\pm0.011$  & $3410\pm180$ & $1.03\pm0.005$   &$0.000015\pm0.000006$& 3/8\\
                 & 50--70\%         & --         &$0.121\pm0.007$  & $0.304\pm0.010$  & $3350\pm200$ & $1.04\pm0.005$   &$0.000007\pm0.0000005$& 11/8\\
\hline
                 & 0--10\%          & $p$       &$0.105\pm0.007$  & $0.309\pm0.008$  & $3130\pm165$ & $1.03\pm0.003$   &$2.4\times10^{-7}\pm4\times10^{-8}$ & 52/9\\
                 & 10--30\%         & --        &$0.104\pm0.006$ & $0.277\pm0.009$ & $2945\pm150$ & $1.034\pm0.005$  &$1.6\times10^{-7}\pm5\times10^{-8}$ & 22/9\\
                 & 30--50\%         &--         &$0.102\pm0.007$  & $0.247\pm0.011$  & $2870\pm135$ & $1.037\pm0.006$  &$6.6\times10^{-8}\pm7\times10^{-9}$ & 16/9\\
                 & 50--70\%         & --        &$0.100\pm0.008$  & $0.227\pm0.010$  & $2797\pm180$ & $1.04\pm0.005$ &$2.3\times10^{-8}\pm5\times10^{-9}$ & 20/9\\
\hline
                 & 0--10\%          & $K^0_S$   &$0.128\pm0.007$  & $0.361\pm0.008$  & $3550\pm192$ & $1.05\pm0.004$   &$0.02\pm0.004$   & 16/12\\
Fig. 2           & 10--20\%         & --        &$0.127\pm0.005$  & $0.342\pm0.009$  & $3465\pm170$ & $1.055\pm0.005$  &$0.0014\pm0.0005$& 5/12\\
Cu-Cu            & 20--30\%         &--         &$0.125\pm0.005$  & $0.334\pm0.009$  & $3387\pm170$ & $1.06\pm0.004$   &$9\times10^{-5}\pm4\times10^{-6}$& 4/12\\
200 GeV          & 30--40\%         & --        &$0.123\pm0.006$  & $0.319\pm0.010$  & $3262\pm164$ & $1.064\pm0.004$  &$6\times10^{-6}\pm5\times10^{-7}$ & 4/12\\
                 & 40--60\%         & --        &$0.120\pm0.007$  & $0.306\pm0.011$  & $3100\pm129$ & $1.069\pm0.005$ &$3.6\times10^{-7}\pm7\times10^{-8}$ & 8/12\\
\hline
                 & 0--10\%          & $\Lambda$ &$0.129\pm0.009$  & $0.303\pm0.010$  & $2990\pm160$ & $1.045\pm0.005$  &$0.006\pm0.0007$ & 7/11\\
                 & 10--20\%         & --       &$0.127\pm0.006$  & $0.286\pm0.006$  & $2800\pm155$ & $1.048\pm0.006$  &$4.1\times10^{-4}\pm2\times10^{-5}$  & 11/11\\
                 & 20--30\%         &--        &$0.124\pm0.004$  & $0.276\pm0.011$  & $2710\pm170$ & $1.05\pm0.004$   &$3\times10^{-5}\pm4\times10^{-6}$ & 2/11\\
                 & 30--40\%         & --       &$0.123\pm0.006$  & $0.267\pm0.007$  & $2587\pm150$ & $1.053\pm0.004$  &$2\times10^{-6}\pm4\times10^{-7}$& 2/11\\
                 & 40--60\%         & --       &$0.120\pm0.007$  & $0.255\pm0.006$  & $2490\pm160$ & $1.055\pm0.005$&$1\times10^{-7}\pm6\times10^{-8}$    & 3/11\\
\hline
                 & 0--10\%          & $\Xi$    &$0.143\pm0.005$  & $0.290\pm0.007$  & $2770\pm220$ & $1.037\pm0.005$  &$0.0009\pm0.00007$& 2/6\\
                 & 10--20\%         & --       &$0.142\pm0.004$  & $0.281\pm0.008$  & $2645\pm200$ & $1.043\pm0.004$  &$4.8\times10^{-5}\pm5\times10^{-6}$   & 5/6\\
                 & 20--30\%         &--        &$0.137\pm0.005$  & $0.270\pm0.009$  & $2500\pm190$ & $1.048\pm0.004$  &$3.4\times10^{-6}\pm8\times10^{-7}$   & 3/6\\
                 & 30--40\%         & --       &$0.136\pm0.006$  & $0.258\pm0.008$  & $2400\pm205$ & $1.052\pm0.004$  &$2.2\times10^{-7}\pm7\times10^{-8}$   & 3/6\\
                 & 40--60\%         & --      &$0.134\pm0.005$  & $0.250\pm0.012$  & $2300\pm210$ & $1.053\pm0.003$  &$1.1\times10^{-8}\pm7\times10^{-9}$ & 3/6\\
\hline
                 & 0--10\%          & $\Omega+\bar\Omega$& $0.144\pm0.005$  & $0.278\pm0.008$ & $2400\pm180$ & $1.055\pm0.004$  &$0.0001\pm0.00005$& 0.6/1\\
                 & 10--20\%         & --  & $0.142\pm0.006$  & $0.270\pm0.010$  & $2313\pm200$ & $1.057\pm0.005$  &$6.7\times10^{-6}\pm7\times10^{-7}$ & 0.5/1\\
                 & 20--30\%         &--    &$0.140\pm0.005$  & $0.260\pm0.008$  & $2200\pm200$ & $1.059\pm0.004$  &$4.3\times10^{-7}\pm6\times10^{-8}$  & 1/1\\
                 & 30--40\%         & --   &$0.138\pm0.006$  & $0.238\pm0.005$  & $2100\pm200$ & $1.060\pm0.004$  &$3\times10^{-8}\pm4\times10^{-9}$  & 0.5/1\\
                 & 40--60\%         & --   &$0.136\pm0.005$  & $0.226\pm0.010$  & $2000\pm170$ & $1.062\pm0.003$  &$1.2\times10^{-9}\pm8\times10^{-10}$  & 0.3/1\\
\hline
Fig. 3           & 0--5\%           & $\pi^+$     &$0.112\pm0.008$  & $0.369\pm0.014$  & $5400\pm300$ & $1.01\pm0.006$   &$0.24\pm0.007$    & 7/6\\
Au-Au            & 5--10\%          & --          &$0.110\pm0.007$  & $0.359\pm0.014$  & $5200\pm400$ & $1.012\pm0.008$  &$0.19\pm0.005$    & 8/6\\
62.4 GeV         & 10--20\%         &--           &$0.108\pm0.008$  & $0.344\pm0.015$  & $5125\pm240$ & $1.014\pm0.009$  &$0.15\pm0.006$    & 5/6\\
                 & 20--30\%         & --          &$0.105\pm0.005$  & $0.331\pm0.014$  & $5000\pm200$ & $1.016\pm0.008$  &$0.11\pm0.008$    & 5/6\\
                 & 30--40\%         & --          &$0.102\pm0.010$  & $0.320\pm0.016$  & $4850\pm190$ & $1.019\pm0.008$  &$0.077\pm0.006$   & 3/6\\
                 & 40--50\%         & --          &$0.100\pm0.009$  & $0.306\pm0.015$  & $4700\pm225$ & $1.022\pm0.01$   &$0.053\pm0.004$   & 11/6\\
                 & 50--60\%         & --          &$0.097\pm0.007$  & $0.290\pm0.013$  & $4600\pm200$ & $1.024\pm0.01$   &$0.034\pm0.005$   & 19/6\\
                 & 60--70\%         & --         &$0.095\pm0.008$  & $0.270\pm0.015$  & $4430\pm370$ & $1.027\pm0.007$  &$0.019\pm0.003$   & 13/6\\
                 & 70--80\%         & --         &$0.092\pm0.008$  & $0.252\pm0.016$  & $4260\pm280$ & $1.03\pm0.014$   &$0.010\pm0.002$   & 15/6\\
\hline
                 & 0--5\%           & $K^+$      &$0.135\pm0.009$  & $0.352\pm0.014$  & $4950\pm320$ & $1.04\pm0.007$   &$0.046\pm0.004$   & 3/6\\
                 & 5--10\%          & --         &$0.133\pm0.008$  & $0.338\pm0.012$  & $4830\pm240$ & $1.043\pm0.008$  &$0.04\pm0.005$    & 0.4/6\\
                 & 10--20\%         &--          &$0.132\pm0.006$  & $0.325\pm0.013$  & $4700\pm220$ & $1.045\pm0.006$  &$0.03\pm0.005$    & 2/6\\
                 & 20--30\%         & --         &$0.130\pm0.008$  & $0.311\pm0.012$  & $4600\pm200$ & $1.048\pm0.007$  &$0.021\pm0.003$   & 1/6\\
                 & 30--40\%         & --         &$0.127\pm0.006$  & $0.300\pm0.013$  & $4460\pm200$ & $1.05\pm0.01$    &$0.014\pm0.004$   & 2/6\\
                 & 40--50\%         & --         &$0.125\pm0.008$  & $0.289\pm0.014$  & $4330\pm200$ & $1.052\pm0.01$   &$0.0095\pm0.0005$ & 6/6\\
                 & 50--60\%         & --         &$0.123\pm0.007$  & $0.275\pm0.014$  & $4200\pm200$ & $1.055\pm0.01$   &$0.0055\pm0.0007$ & 0.4/6\\
                 & 60--70\%         & --         &$0.120\pm0.008$  & $0.261\pm0.013$  & $4076\pm250$ & $1.058\pm0.009$  &$0.0031\pm0.0004$ & 5/6\\
                 & 70--80\%         & --         &$0.117\pm0.007$  & $0.237\pm0.011$  & $3900\pm220$ & $1.06\pm0.01$    &$0.0012\pm0.0003$ & 6/6\\
\hline
                 & 0--5\%           & $p$       &$0.113\pm0.008$  & $0.336\pm0.012$  & $4300\pm200$ & $1.1\pm0.013$     &$0.024\pm0.005$   & 1/10\\
                 & 5--10\%          & --        &$0.110\pm0.009$  & $0.325\pm0.013$  & $4200\pm230$ & $1.13\pm0.01$     &$0.0215\pm0.004$  & 5/10\\
                 & 10--20\%         &--         &$0.108\pm0.007$  & $0.318\pm0.013$  & $4040\pm200$ & $1.14\pm0.008$    &$0.018\pm0.006$   & 5/10\\
                 & 20--30\%         & --        &$0.106\pm0.008$  & $0.307\pm0.015$  & $3900\pm270$ & $1.15\pm0.01$     &$0.0125\pm0.004$  & 11/10\\
                 & 30--40\%         & --        &$0.104\pm0.008$  & $0.300\pm0.014$  & $3770\pm185$ & $1.16\pm0.008$    &$0.0085\pm0.0007$ & 13/10\\
                 & 40--50\%         & --        &$0.101\pm0.009$  & $0.288\pm0.013$  & $3600\pm220$ & $1.17\pm0.008$    &$0.005\pm0.0005$  & 19/10\\
                 & 50--60\%         & --        &$0.099\pm0.007$  & $0.276\pm0.011$  & $3500\pm180$ & $1.18\pm0.005$   &$0.003\pm0.0002$   & 13/10\\
                 & 60--70\%         & --        &$0.097\pm0.008$  & $0.260\pm0.012$  & $3400\pm180$ & $1.19\pm0.005$   &$0.0015\pm0.0004$  & 12/10\\
                 & 70--80\%         & --        &$0.095\pm0.009$  & $0.246\pm0.014$  & $3330\pm200$ & $1.2\pm0.005$&$6.5\times10^{-4}\pm7\times10^{-5}$ & 4/10\\
\hline
                 & 0--5\%           & $K^0_S$   &$0.135\pm0.007$  & $0.347\pm0.007$  & $4900\pm300$ & $1.034\pm0.005$   &$0.023\pm0.005$    & 1/10\\
Fig. 4           & 5--10\%          & --        &$0.132\pm0.005$  & $0.338\pm0.007$  & $4800\pm240$ & $1.038\pm0.004$  &$0.002\pm0.0004$    & 2/10\\
Au-Au            & 10--20\%         &--         &$0.130\pm0.007$  & $0.322\pm0.008$  & $4670\pm360$ & $1.045\pm0.004$  &$0.00016\pm0.00005$ & 1/10\\
62.4 GeV         & 20--30\%         & --       &$0.125\pm0.006$  & $0.309\pm0.009$  & $4500\pm245$ & $1.047\pm0.005$  &$1.1\times10^{-5}\pm4\times10^{-6}$ & 2/10\\
                 & 30--40\%         & --       &$0.122\pm0.006$  & $0.296\pm0.007$  & $4380\pm300$ & $1.053\pm0.005$  &$1\times10^{-6}\pm3\times10^{-7}$   & 1/10\\
                 & 40--60\%         & --       &$0.120\pm0.006$  & $0.282\pm0.009$  & $4200\pm215$ & $1.056\pm0.006$  &$4.5\times10^{-8}\pm4\times10^{-9}$ & 2/10\\
                 & 60--80\%         & --       &$0.118\pm0.007$  & $0.274\pm0.010$  & $4100\pm180$ & $1.058\pm0.005$  &$1.6\times10^{-9}\pm6\times10^{-10}$ & 3/10\\
\hline
                 & 0--5\%           & $\Lambda$ &$0.136\pm0.006$  & $0.224\pm0.010$  & $3790\pm260$ & $1.036\pm0.004$  &$0.02\pm0.004$    & 4/8\\
                 & 5--10\%          & --        &$0.135\pm0.007$  & $0.215\pm0.012$  & $3600\pm300$ & $1.037\pm0.004$   &$1.6\times10^{-4}\pm4\times10^{-5}$  & 9/8\\
                 & 10--20\%         &--         &$0.133\pm0.005$  & $0.205\pm0.011$  & $3470\pm185$ & $1.039\pm0.005$  &$1.4\times10^{-4}\pm5\times10^{-5}$ & 7/8\\
                 & 20--30\%         & --        &$0.132\pm0.006$  & $0.195\pm0.010$  & $3356\pm200$ & $1.040\pm0.005$  &$8\times10^{-6}\pm5\times10^{-7}$    & 11/8\\
                 & 30--40\%         & --        &$0.130\pm0.005$  & $0.185\pm0.011$  & $3210\pm170$ & $1.042\pm0.006$  &$7\times10^{-7}\pm4\times10^{-8}$    & 3/8\\
\hline
\end{tabular}%
\end{center}}
\end{table*}
\begin{table*}
{\scriptsize To be continued. \vspace{-.50cm}
\begin{center}
\begin{tabular}{ccccccccccc}\\ \hline\hline
                 & 40--60\%         & --   &$0.128\pm0.007$  & $0.172\pm0.013$  & $3100\pm190$ & $1.044\pm0.005$  &$2.7\times10^{-8}\pm6\times10^{-9}$  & 19/8\\
                 & 60--80\%         & --   &$0.125\pm0.006$  & $0.165\pm0.010$  & $3000\pm170$ & $1.050\pm0.004$  &$7\times10^{-10}\pm8\times10^{-11}$   & 5/8\\
   \hline
               & 0--5\%           & $\bar\Xi^+$ &$0.151\pm0.006$  & $0.215\pm0.007$  & $3000\pm260$ & $1.015\pm0.004$   &$0.0014\pm0.0003$& 1/6\\
                 & 5--10\%         & --        &$0.149\pm0.007$  & $0.206\pm0.009$  & $2800\pm200$ & $1.02\pm0.004$    &$0.00012\pm0.00005$ & 3/7\\
                 & 10--20\%         &-- &$0.147\pm0.005$  & $0.197\pm0.010$  & $2680\pm156$ & $1.024\pm0.004$  &$8.5\times10^{-7}\pm4\times10^{-8}$   & 1/7\\
                 & 20--40\%         & -- &$0.145\pm0.006$  & $0.184\pm0.009$  & $2600\pm140$ & $1.025\pm0.005$  &$5.5\times10^{-7}\pm3\times10^{-8}$& 2/7\\
                 & 40--60\%         & --  &$0.143\pm0.005$  & $0.180\pm0.008$  & $2500\pm144$ & $1.028\pm0.005$  &$1.6\times10^{-8}\pm6\times10^{-9}$& 2/6\\
                 & 60--80\%         & --  &$0.141\pm0.006$  & $0.170\pm0.010$  & $2400\pm150$ & $1.029\pm0.004$  &$4.8\times10^{-10}\pm7\times10^{-11}$ & 4/5\\
\hline
                 & 0--20\%          & $\bar\Omega^+$ &$0.153\pm0.006$  & $0.170\pm0.008$ & $2400\pm200$ & $1.025\pm0.004$  &$0.0001\pm0.00003$  & 1/1\\
                 & 20--40\%         & -- &$0.150\pm0.007$  & $0.160\pm0.012$  & $2260\pm170$ & $1.035\pm0.005$  &$3\times10^{-7}\pm7\times10^{-8}$  & 0.4/1\\
                 & 40--60\%         &--  &$0.148\pm0.005$  & $0.150\pm0.011$  & $2160\pm138$ & $1.055\pm0.008$   &$4\times10^{-9}\pm6\times10^{-10}$& 1/0\\
\hline
Fig. 5           & 0--5\%           & $\pi^+$  &$0.126\pm0.005$  & $0.440\pm0.013$  & $8660\pm400$ & $1.02\pm0.006$   &$290\pm45$    & 215/37\\
Pb-Pb            & 5--10\%          & --       &$0.124\pm0.005$  & $0.430\pm0.011$  & $8530\pm370$ & $1.029\pm0.006$  &$120\pm27$    & 191/37\\
2.76 TeV         & 10--20\%         &--        &$0.122\pm0.005$  & $0.416\pm0.012$  & $8400\pm290$ & $1.034\pm0.005$  &$50\pm7$      & 211/37\\
                 & 20--30\%         & --       &$0.120\pm0.005$  & $0.407\pm0.009$  & $8260\pm300$ & $1.042\pm0.005$  &$17\pm4$      & 203/37\\
                 & 30--40\%         & --       &$0.117\pm0.006$  & $0.399\pm0.012$  & $8000\pm360$ & $1.046\pm0.004$  &$6\pm1$       & 274/37\\
                 & 40--50\%         & --       &$0.115\pm0.007$  & $0.389\pm0.010$  & $7800\pm330$ & $1.054\pm0.006$  &$1.8\pm0.3$   & 186/37\\
                 & 50--60\%         & --       &$0.113\pm0.006$  & $0.378\pm0.011$  & $7620\pm300$ & $1.059\pm0.005$  &$0.5\pm0.03$  & 181/37\\
                 & 60--70\%         & --       &$0.110\pm0.005$  & $0.364\pm0.010$  & $7500\pm290$ & $1.062\pm0.004$  &$0.13\pm0.04$ & 339/37\\
                 & 70--80\%         & --       &$0.108\pm0.005$  & $0.350\pm0.012$  & $7280\pm280$ & $1.069\pm0.005$  &$0.03\pm0.002$& 255/37\\
                 & 80--90\%         & --       &$0.106\pm0.006$  & $0.340\pm0.011$  & $7140\pm210$ & $1.073\pm0.006$  &$0.006\pm0.0005$& 255/37\\
\hline
                 & 0--5\%           & $K^+$    &$0.142\pm0.007$  & $0.422\pm0.009$  & $8100\pm300$ & $1.045\pm0.004$   &$40\pm7$        & 21/32\\
                 & 5--10\%          & --       &$0.140\pm0.006$  & $0.410\pm0.013$  & $7900\pm320$ & $1.047\pm0.005$   &$20\pm4$         & 32/32\\
                 & 10--20\%         &--        &$0.136\pm0.005$  & $0.396\pm0.011$  & $7760\pm270$ & $1.058\pm0.008$    &$7\pm0.2$        & 29/32\\
                 & 20--30\%         & --       &$0.133\pm0.005$  & $0.382\pm0.012$  & $7600\pm220$ & $1.062\pm0.005$   &$2.5\pm0.5$      & 40/32\\
                 & 30--40\%         & --      &$0.130\pm0.007$  & $0.375\pm0.010$  & $7500\pm200$ & $1.07\pm0.006$    &$0.8\pm0.03$     & 23/32\\
                 & 40--50\%         & --      &$0.128\pm0.005$  & $0.361\pm0.013$  & $7350\pm230$ & $1.076\pm0.006$  &$0.25\pm0.04$    & 10/32\\
                 & 50--60\%         & --      &$0.125\pm0.006$  & $0.350\pm0.011$  & $7220\pm210$ & $1.08\pm0.005$   &$0.07\pm0.004$   & 13/32\\
                 & 60--70\%         & --      &$0.124\pm0.006$  & $0.340\pm0.010$  & $7100\pm180$ & $1.085\pm0.006$  &$0.018\pm0.003$  & 104/32\\
                 & 70--80\%         & --    &$0.122\pm0.005$  & $0.327\pm0.012$  & $6940\pm220$ & $1.089\pm0.005$  &$0.004\pm0.0005$ & 15/32\\
                 & 80--90\%         & --     &$0.120\pm0.007$  & $0.317\pm0.010$  & $6810\pm265$ & $1.09\pm0.006$   &$0.00082\pm0.00004$& 11/32\\
\hline
                 & 0--5\%           & $p$      &$0.127\pm0.007$  & $0.410\pm0.009$  & $7467\pm260$ & $1.065\pm0.005$   &$9.82\pm0.6$     & 316/33\\
                 & 5--10\%          & --       &$0.125\pm0.006$  & $0.400\pm0.011$  & $7300\pm300$ & $1.07\pm0.005$    &$3.79\pm0.4$     & 335/33\\
                 & 10--20\%         &--        &$0.123\pm0.005$  & $0.388\pm0.012$  & $7170\pm190$ & $1.078\pm0.007$    &$1.49\pm0.2$     & 273/33\\
                 & 20--30\%         & --       &$0.121\pm0.006$  & $0.373\pm0.013$  & $7000\pm215$ & $1.082\pm0.004$   &$0.49\pm0.05$    & 223/33\\
                 & 30--40\%         & --       &$0.119\pm0.005$  & $0.361\pm0.009$  & $6800\pm200$ & $1.084\pm0.004$   &$0.17\pm0.03$    & 168/33\\
                 & 40--50\%         & --       &$0.116\pm0.007$  & $0.348\pm0.011$  & $6692\pm192$ & $1.087\pm0.006$   &$0.052\pm0.004$  & 116/33\\
                 & 50--60\%         & --       &$0.113\pm0.006$  & $0.338\pm0.010$  & $6520\pm175$ & $1.088\pm0.004$   &$0.0145\pm0.004$ & 35/33\\
                 & 60--70\%         & --       &$0.110\pm0.006$  & $0.325\pm0.012$  & $6400\pm200$ & $1.09\pm0.004$    &$0.004\pm0.0004$ & 19/33\\
                 & 70--80\%         & --       &$0.108\pm0.007$  & $0.311\pm0.011$  & $6300\pm160$ & $1.092\pm0.005$   &$0.00084\pm0.00004$& 24/33\\
                 & 80--90\%         & --       &$0.105\pm0.005$  & $0.291\pm0.013$  & $6250\pm250$ & $1.094\pm0.004$   &$0.00017\pm0.00006$& 113/33\\
\hline
                 & 0--10\%          & $K^0_S$    &$0.142\pm0.007$  & $0.420\pm0.015$  & $8020\pm380$ & $1.045\pm0.004$  &$0.47\pm0.07$    & 33/17\\
Fig. 6           & 10--20\%         & --         &$0.140\pm0.006$  & $0.410\pm0.012$  & $7900\pm295$ & $1.05\pm0.005$    &$0.35\pm0.06$   & 16/17\\
Pb-Pb            & 20--40\%         &--          &$0.138\pm0.006$  & $0.400\pm0.010$  & $7770\pm257$ & $1.057\pm0.005$   &$0.20\pm0.04$   & 13/17\\
2.76 TeV         & 40--60\%         & --         &$0.136\pm0.008$  & $0.387\pm0.013$  & $7640\pm245$ & $1.062\pm0.006$   &$0.072\pm0.003$ & 16/17\\
                 & 60--80\%         & --         &$0.133\pm0.006$  & $0.374\pm0.010$  & $7510\pm215$ & $1.069\pm0.004$   &$0.019\pm0.003$ & 120/17\\
\hline
                 & 0--10\%          & $\Lambda$  &$0.143\pm0.006$  & $0.400\pm0.011$  & $7010\pm335$ & $1.033\pm0.004$   &$0.055\pm0.005$   & 39/15\\
                 & 10--20\%         & --         &$0.141\pm0.005$  & $0.389\pm0.010$  & $6900\pm270$ & $1.035\pm0.005$   &$0.042\pm0.0055$  & 15/15\\
                 & 20--40\%         &--          &$0.138\pm0.005$  & $0.375\pm0.009$  & $6770\pm256$ & $1.038\pm0.004$   &$0.026\pm0.004$  & 10/15\\
                 & 40--60\%         & --         &$0.136\pm0.007$  & $0.355\pm0.012$  & $6660\pm240$ & $1.039\pm0.005$   &$0.01\pm0.002$  & 52/15\\
                 & 60--80\%         & --         &$0.133\pm0.006$  & $0.302\pm0.015$  & $6560\pm195$ & $1.041\pm0.004$   &$0.00278\pm0.00006$& 68/15\\
\hline
                 & 0--10\%          & $\Xi$      &$0.157\pm0.008$  & $0.385\pm0.015$  & $6560\pm260$ & $1.03\pm0.004$   &$0.0155\pm0.003$   & 8/8\\
                 & 10--20\%         & --         &$0.155\pm0.006$  & $0.375\pm0.013$  & $6460\pm240$ & $1.032\pm0.005$  &$0.0118\pm0.004$   & 17/8\\
                 & 20--40\%         &--          &$0.153\pm0.007$  & $0.342\pm0.010$  & $6320\pm320$ & $1.034\pm0.006$  &$0.0075\pm0.00004$ & 18/8\\
                 & 40--60\%         & --        &$0.150\pm0.005$  & $0.302\pm0.012$  & $6200\pm210$ & $1.036\pm0.006$  &$0.003\pm0.0002$   & 17/8\\
                 & 60--80\%         & --        &$0.146\pm0.008$  & $0.287\pm0.009$  & $6154\pm194$ & $1.04\pm0.005$  &$0.00065\pm0.00005$ & 39/8\\
\hline
                 & 0--10\%          & $\Omega$   &$0.158\pm0.007$  & $0.350\pm0.013$ & $6120\pm175$ & $1.04\pm0.004$    &$0.0012\pm0.0003$    & 0.3/3\\
                 & 10--20\%         & --         &$0.155\pm0.005$  & $0.330\pm0.010$  & $6000\pm220$ & $1.054\pm0.005$  &$0.0008\pm0.00004$   & 1/3\\
                 & 20--40\%         &--         &$0.153\pm0.006$  & $0.280\pm0.009$  & $5800\pm200$ & $1.06\pm0.006$   &$0.0005\pm0.00003$   & 3/3\\
                 & 40--60\%         & --       &$0.149\pm0.006$  & $0.220\pm0.014$  & $5700\pm240$ & $1.062\pm0.008$   &$0.0002\pm0.00003$  & 7/3\\
                 & 60--80\%         & --       &$0.147\pm0.005$  & $0.190\pm0.011$  & $5500\pm214$ & $1.065\pm0.007$  &$3.5\times10^{-5}\pm5\times10^{-6}$ &3/2\\
\hline
\end{tabular}%
\end{center}}
\end{table*}

Figure 5 and 6 are the same as Figure 1, but it shows the $p_T$ spectra of
$\pi^+$, $K^+$, $p$ and $K^0_S$ in panels of fig5.(a)-5(d)  and $\Lambda$, $\Xi$ and
$\Omega$ in fig6.(a)-6(c) in Pb-Pb collisions at 2.76 TeV in
$|y|<0.5$. The spectra is distributed in different centrality classes,
e.g for $\pi^+$, $K^+$ and $p$ 0--5\%, 5--10\%, 10--20\%, 20--30\%, 30--40\%, 40--50\%, 50--60\%, 60--70\%, 70--80\% and 80--90\% , while for
$K^0_S$, $\Lambda$, $\Xi$ and $\Omega$ 0--10\%, 10--20\%, 20--40\%, 40--60\% and 60--80\%. The symbols
represent the experimental data of ALICE Collaboration [47, 48] (the data for $\pi^+$, $K^+$ and $p$ is taken from ref.47 and data for $K^0_S$, $\Lambda$, $\Xi$ and $\Omega$ is taken from 48), while the curves are the results of
our fitting by using Eq. (1). Each panel is followed by result of its
data/fit. One can see that the model results describe the experimental data in special $p_T$ ranges at LHC. We comment here by the way that the description of the pion spectra at LHC is poor due to the non-inclusion of the resonance generation in low $p_T$ region.
In Figs. 1-6, $p_T$ spectra of some particles are re-scaled in some centrality bins which are showed in the corresponding figures.
\begin{figure*}[htb!]
\begin{center}
\hskip-0.153cm
\includegraphics[width=15cm]{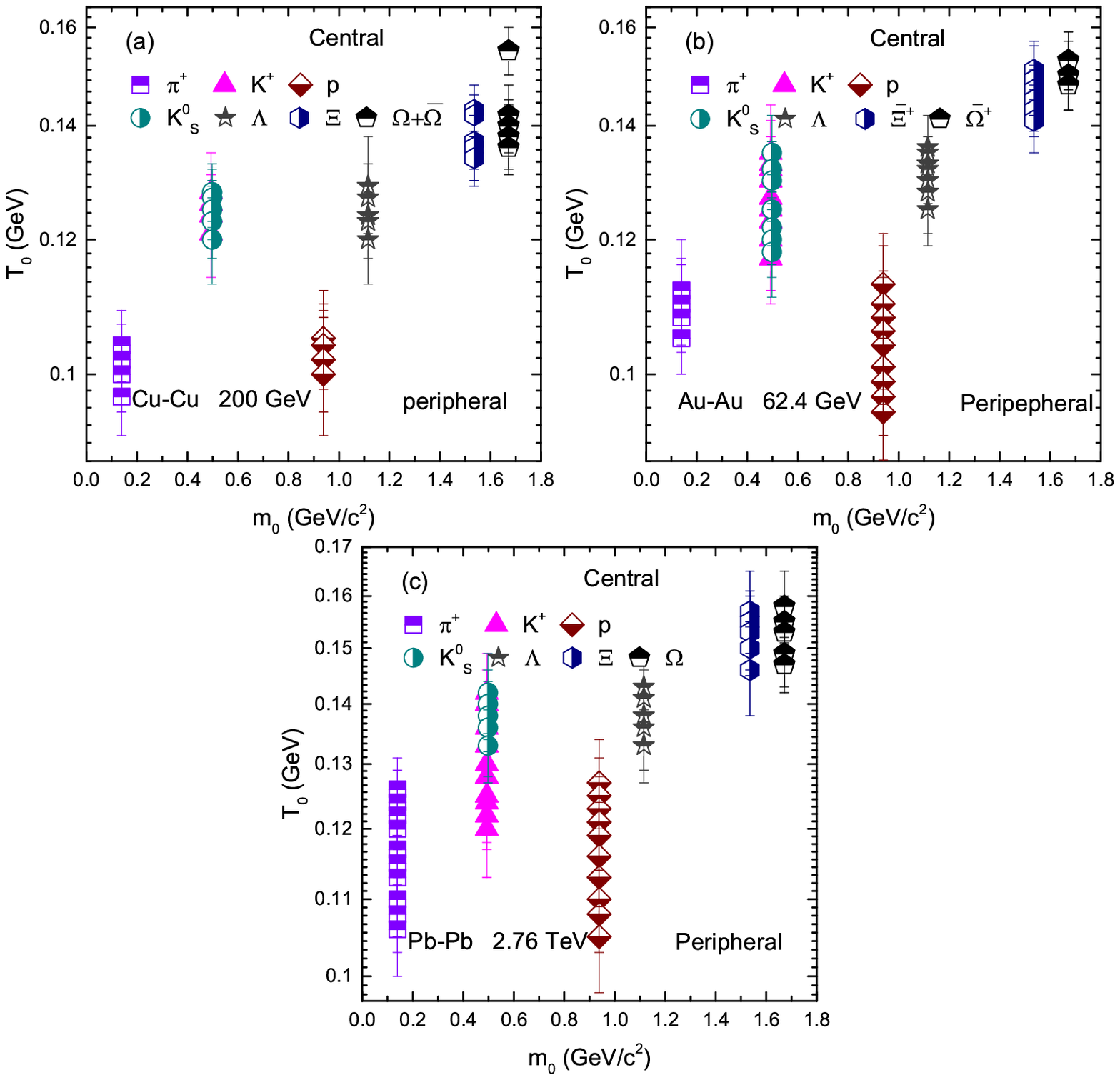}
\end{center}
Fig. 7. Dependence of kinetic freezeout temperature on the rest mass and centrality.
\end{figure*}

To study the changing tendencies of parameters, figure 7 shows the dependence of $T_0$ on the rest mass of the particle as well as on centrality. Different symbols represent different particles and the trend of symbols from up to downwards show the trend from central towards peripheral collisions in panels (a)-(c). One can see that $T_0$ of the emission source extracted in different centrality intervals is the same for $\pi^+$ and $p$, similarly it is the same for $K^+$, $K^0_S$ and $\Lambda$, and $\Xi$ or $\bar \Xi^+$ and $\Omega+\bar\Omega$ or $\bar \Omega^+$ or $\Omega$. In the analysed particles, $\Xi$ or $\bar \Xi^+$ and $\Omega+\bar\Omega$ or $\bar \Omega^+$ or $\Omega$ are the heaviest particles and they freezeout earlier than the rest of others. $\pi^+$ is the lightest particle and it freezeout after $\Xi$ or $\bar \Xi^+$, $\Omega+\bar\Omega$ or $\bar \Omega^+$ or $\Omega$, $K^+$, $K^0_S$ and $\Lambda$. $p$ is heavier than $\pi$, $K^+$ and $K^0_S$ but it freezeout after $K^+$ and $K^0_S$ (same result of larger $T_0$ for $K^+$ and $K^0_S$ than $p$ can be found in [49], although the main idea is different from our present work). Eventually, the freezeout of $p$ occurs at the same time with $\pi$. At present, there is no clear dependence of $T_0$ on mass of the particle. Therefore we believe that the recent result of $T_0$ of the particles may be dependent on the production cross-section of the interacting particle, such that, smaller the  production cross section of the interacting particle, larger will be its source $T_0$ and the particle will freezeout earlier. $\pi$ and $p$ are non-strange particles and have larger production cross-section, while $K^+$, $K^0_S$ and $\Lambda$ are strange and  $\Xi$ or $\bar \Xi^+$ and $\Omega+\bar\Omega$ or $\bar \Omega^+$ or $\Omega$ are the multi-strange particles and the later has the smallest production cross-section and they freezeout earlier than the strange particles. The above statement reveals the scenario of triple kinetic freezeout due to the separate decoupling of the non-strange, strange and multi-strange particles and this scenario is observed by us (single, double and multi-kinetic freezeout scenarios are already studied in different literatures). Furthermore, $T_0$ depends on centrality and it shows a decreasing trend as we go from central to peripheral collisions. In central collision, large number of particles get involved in interactions due to large participant collision cross-section of the interacting system and deposits more energy in the system which results in larger kinetic freezeout temperature due to higher degree of excitation of the system [14, 15, 50]. However the cross section of the interacting system decrease with decreasing centrality that leads the decrease of kinetic freezeout temperature. The kinetic freezeout temperature is observed to be larger in Pb-Pb than in Au-Au collisions and in the later case it is larger than Cu-Cu collisions.
\begin{figure*}[htb!]
\begin{center}
\hskip-0.153cm
\includegraphics[width=15cm]{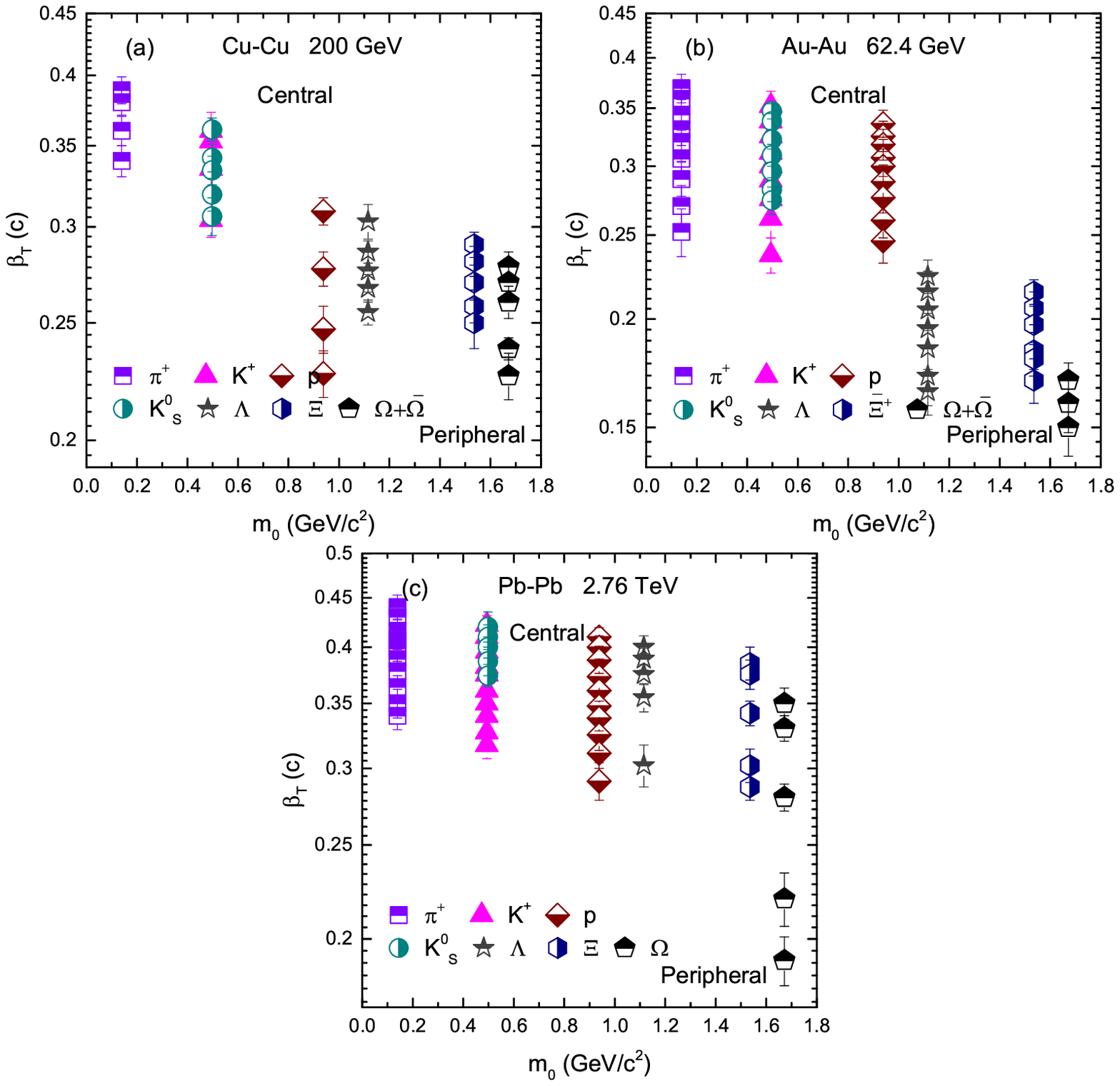}
\end{center}
Fig. 8. Shows the dependence of transverse flow velocity on rest mass of the particle and centrality.
\end{figure*}

Figure 8 is similar to Figure 7, but it represents the dependence of $\beta_T$ on the rest mass of the particle and on centrality. One can see that greater the mass of the particle is, smaller the value for $\beta_T$. It is also observed that $\beta_T$ is decreasing with decreasing the event centrality due to decreasing the participant collision cross-section of the interacting system. In larger cross-section interacting system, large amount of energy is stored due to the participation of large number of nucleons in the interaction and thus the system undergoes a rapid expansion but this expansion becomes more and more steady from center to periphery. $\beta_T$ is found to be slightly larger in Pb-Pb than in Cu-Cu and in the later it is larger than in Au-Au which shows its dependence on the center of mass energy of the system.
\begin{figure*}[htb!]
\begin{center}
\hskip-0.153cm
\includegraphics[width=15cm]{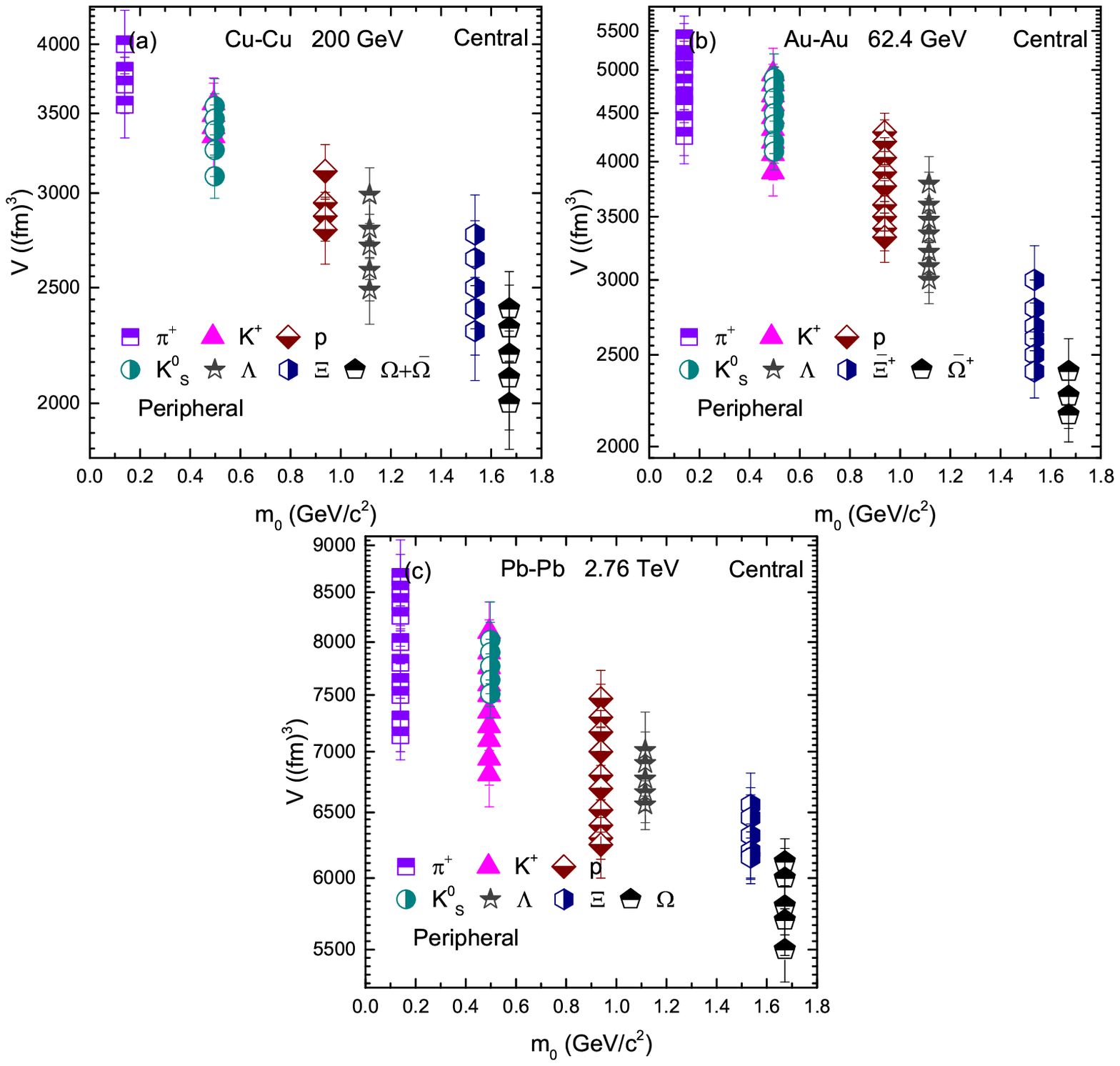}
\end{center}
Fig. 9. Shows dependence of the kinetic freezeout volume $V$ on the rest mass $m_0$ and centrality.
\end{figure*}

Figure 9 is also similar to Figure 7, but it exhibits the dependence of kinetic freezeout volume on rest mass of and centrality. The volume differential scenario is observed as the the kinetic freezeout volume decreases for heavier particles, which shows the early freezeout of the heavy particles than the lighter ones and this maybe the tip-off of different freezeout surfaces for different particles. The kinetic freezeout volume is observed to be larger in Pb-Pb than in Au-Au collisions and in the later it is larger than Cu-Cu collisions. The kinetic freezeout volume has a decreasing trend from central to peripheral collisions due to decreasing the number of participant nucleons from central to periphery. Because the mass of K$^+$ and K$^0_\mathrm{S}$ is almost identical and the relative parameter values give very close values to each other, so the K$^+$ data points are  hidden beneath  K$^0_\mathrm{S}$.
\begin{figure*}[htb!]
\begin{center}
\hskip-0.153cm
\includegraphics[width=15cm]{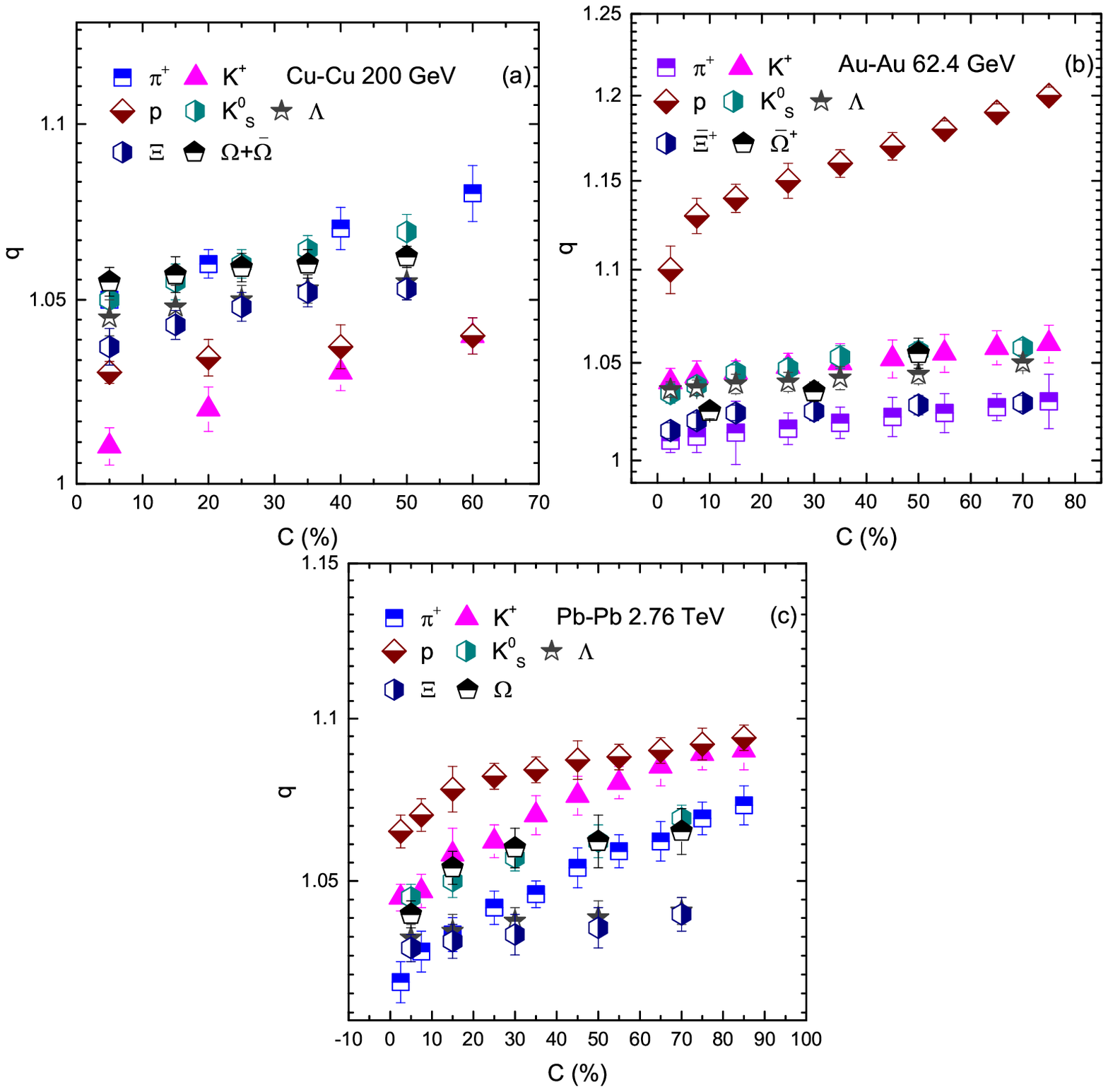}
\end{center}
Fig. 10. Shows the dependence of entropy index q on  centrality.
\end{figure*}
\begin{figure*}[htb!]
\begin{center}
\hskip-0.153cm
\includegraphics[width=15cm]{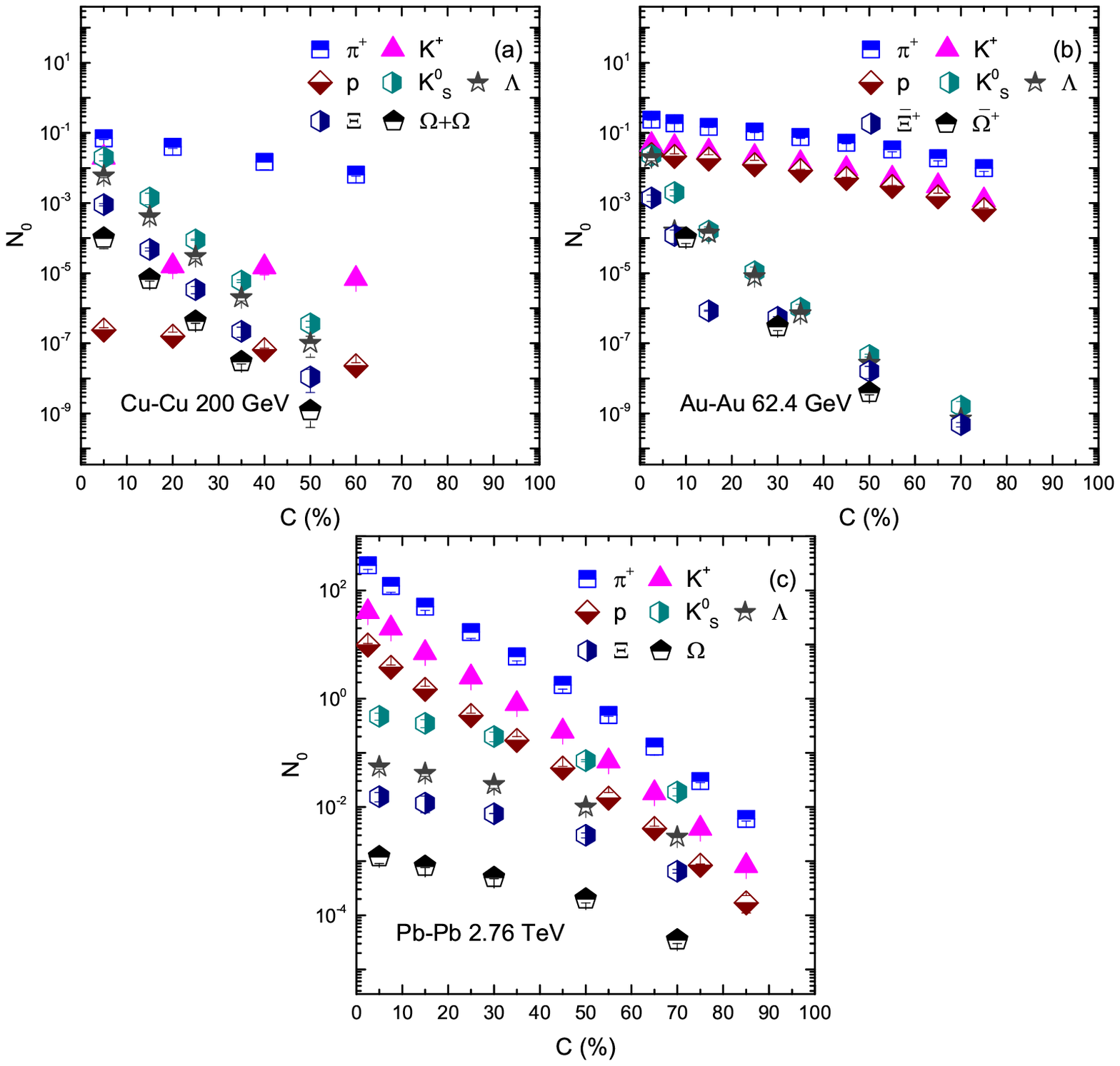}
\end{center}
Fig. 11. Demonstrates the dependence of normalization constant $N_0$ on  centrality.
\end{figure*}

Figure 10 shows the dependence of $q$ on centrality. The more central the collision is, less will be $q$, which indicates to a quick approach of equilibrium in the central collisions. The range of increase of "q" from central collisions to periphery is 0.001 to 0.03. Similarly, Figure 11 shows the dependence of $N_0$ on centrality. $N_0$ has a physical significance and it reflects the multiplicity. $N_0$ decreases with the event centrality which obviously shows lower multiplicity from central to periphery.
\\
 {\section{Conclusions}}
 The main observations and conclusions are summarized here.

 a) The transverse momentum spectra of different particle species are analyzed by the blast wave model with Tsallis statistics and the bulk properties in terms of the kinetic freezeout temperature, transverse flow velocity and kinetic freezeout volume are extracted.

 b) The kinetic freezeout temperature ($T_0$) is observed to be dependent on the cross-section of the interacting particle and interacting system as well. Larger the production cross-section of the interacting particle, smaller will be $T_0$. However the case is reverse for participant collision cross-section of the interacting system where larger cross-section leads to larger $T_0$.

 c) A triple kinetic freezeout scenario is observed due to the separate decoupling of non-strange, strange and strange particles.

 d) The transverse flow velocity ($\beta_T$)  and kinetic freezeout volume ($V$) are observed to be mass dependant. Larger the mass of the particle, smaller the $\beta_T$ and $V$ are. Additionally both of them decrease from central to peripheral collisions.

 e) $V$ is larger in Pb-Pb collisions than in Au-Au collisions and in Au-Au collisions it is larger than Cu-Cu collisions which shows its dependence on  collision energy. $\beta_T$ exhibits the dependence on the center of mass of the interacting system as it is larger in Pb-Pb than in Cu-Cu collisions and in the later case it is larger than in Au-Au collisions. The volume for Pb-Pb collisions is about twice the one for Cu-Cu. This is natural due to the large size for Pb.

f) The entropy index $q$ is observed to increases from central to peripheral collisions while the normalization constant $N_0$ decreases from central to peripheral collisions.
\\

{\bf Data availability}

The data used to support the findings of this study are included
within the article and are cited at relevant places within the
text as references.
\\
\\
{\bf Compliance with Ethical Standards}

The authors declare that they are in compliance with ethical
standards regarding the content of this paper.
\\
\\
{\bf Acknowledgements}

The authors would like to thank support from the National Natural Science
Foundation of China (Grant Nos. 11875052, 11575190, and 11135011).
\\
\\

\end{multicols}
\end{document}